\documentclass[prb,aps]{revtex4}
\usepackage{epsfig,latexsym,amssymb,amsmath,amsbsy}

\begin{document}

\title{Spin Waves in Antiferromagnetic Spin Chains with Long Range 
Interactions}

\author{Eddy Yusuf, Anuvrat Joshi, and Kun Yang}
\affiliation{National High Magnetic Field Laboratory and
Department of Physics, Florida State University, Tallahassee, 
Florida 32306}
\date{\today}
\begin{abstract}
We study antiferromagnetic spin
chains with unfrustrated long-range interactions 
%of the form $J_{ij}\sim (-1)^{i-j+1}|i - j|^{-\beta}$,
that decays as a power law with exponent $\beta$,
using the
spin wave approximation. 
We find for sufficiently large spin 
$S$, the Neel order is stable 
at $T=0$ for $\beta < 3$, and survive up to a finite Neel temperature
for $\beta < 2$, validating the 
spin-wave approach in these regimes. 
We estimate the critical values of $S$ and $T$ for the Neel order to be stable.
The spin wave spectra are found to be gapless but 
have non-linear momentum dependence at long wave length,
which is responsible for the suppression of quantum and 
thermal fluctuations and stabilizing the Neel state. 
We also show that for $\beta\le 1$ and
for a large but finite-size system
size $L$,
the excitation gap of the system approaches
zero slower than $L^{-1}$, a behavior that is in contrast to the
Lieb-Schulz-Mattis theorem. 
\end{abstract}
\maketitle

\section{Introduction}

Antiferromagnetic (AF) spin chains have attracted considerable interest from
physicists
in the last two decades, and continue to be subjects of active research at 
present.
There are several reasons why they are of such strong interest.
Firstly, quantum
antiferromagnetic spin chains are important examples of a larger
class of strongly 
correlated   
systems, 
whose ground state and low-energy behavior differ from their higher 
dimensional counterparts in qualitative ways. 
In the case of AF spin chains, quantum fluctuations destroy the Neel 
order in the ground state no matter how big the the size of the spin is, 
while in higher dimensions the Neel order is stable regardless of spin size,
in the absence of frustration. 
Secondly, the spin chains are of interest to physicists because they are ideal 
playgrounds for various types of theoretical approaches. A prominent example 
here is the work of Haldane \cite{haldane}, who mapped the AF spin chains to 
quantum 
non-linear sigma models, and predicted that the integer chains have a gap in
their
excitation spectra while no gap exists for
half integer chains, based on the absence or presence of a 
topological term in the mapping. This fundamental difference is consistent 
with, and to certain degree implied in the Lieb-Schultz-Mattis
(LSM) theorem \cite{lsm}, which 
states that for a Heisenberg AF chains with length $L$ and 
periodic condition, 
for half-integer spins, there exists
an excited state with energy separated from the ground state that is of order
$1/L$; no such theorem exists for integer chains however.

The studies of AF spin chains, and the results mentioned above, are restricted
to models with short-range interactions. In this work we study AF chains with
interactions that decay as power laws and without frustration:
\begin{equation}
\label{hamiltonian}
H=\sum_{ij} (-1)^{i-j+1}J_{ij}\boldsymbol{S}_i\cdot\boldsymbol{S}_j,
\end{equation}
with
\begin{equation}
\label{powerlaw}
J_{ij} = J/|i-j|^{\beta},
\end{equation}
where $J > 0$ determines the overall
energy scale of the system and $\beta$ is the
power-law exponent that controls the decay of the interaction.
The factor $(-1)^{i-j+1}$ ensures that
spins sitting on 
opposite sublattices have antiferromagnetic interactions and those 
sitting on the same sublattice have ferromagnetic interactions, thus there
is no frustration. 
Our motivation of the study comes from the following 
considerations. Firstly, such power-law long-range interactions can in 
principle be realized in experimental systems; one example of which being the 
RKKY\cite{rkky} interaction
mediated by conduction electrons that decay as power laws, with an exponent  
that depends on the details of the conduction electron Fermi surface.
Secondly, as we will show, such long-range interactions tend to suppress
quantum as well as thermal fluctuations, thus increasing the range of 
interaction has an effect that is somewhat
similar to increasing the 
dimensionality of the system. On the other hand the dimensionality is discrete
while the power-law exponent for the interaction can be tuned continuously, 
thus
providing a tuning parameter for the fluctuations; it is of interest to
study how the system behave under such tuning.

Anticipating the stability of Neel order in the presence of such long-range 
interactions, we study the models using the spin-wave method. We obtain the 
following results. (i) We show the Neel order is stable at zero temperature 
for $\beta < 3$ and sufficiently large $S$, justifying the usage of spin-wave
method in this case. 
We also estimate the critical size of the spin for the Neel order to
be stable, as a function of $\beta$. (ii) In this case 
the spin-wave excitation spectra 
take the form $\omega_k\sim k^\gamma$ in the long
wave-length, with $\gamma < 1$ 
and varying 
continuously with $\beta$. (iii) Extending the spin-wave calculation to finite
temperature, we show that the Neel transition temperature $T_N$ is zero for
$\beta \ge 2$ while finite for $\beta < 2$. We determine $T_N$ as a function
of $S$ and $\beta$.
(iv) For a finite-size system with
size $L$ and periodic boundary condition, and $\beta \le 1$, 
we find the lowest excitation energy 
approaches zero slower than $1/L$ as $L$ increases for both half-integer and
integer spins, thus ``violating"the LSM theorem. Of course 
the LSM theorem applies to spin chains with short-range interaction only; here
we have provided explicit examples of how it is invalidated by the presence of
long-range interaction.  

The remainder of the paper is organized as follows. In Sec. II we 
discuss the application of spin wave technique to this 
model. In Sec.III and IV we present and 
discuss the significance of our results. In Sec. V we 
summarize our work and discuss the implications of our results.

\section{The Spin Wave Approach}

We consider a Heisenberg antiferromagnetic chain with unfrustrated power-law
long-range
interaction with the Hamiltonian given by Eq. (\ref{hamiltonian}).
The central issue we address in this work is the stability of Neel state at
zero or low temperature. It is thus natural to use the spin-wave method based
on the Holstein-Primakoff transformation \cite{hp} that maps spin operators to 
boson operators, and check its self-consistency. The procedure is rather 
standard;\cite{anderson}
we nevertheless include the details here for the sake of 
completeness and establish notation for later treatment.
We divide the chain into two sublattices and represent the spin operators in
terms two types of bosons : $a$ bosons which live on $A$ sublattice and
and $b$ bosons which live on $B$ sublattice. Up to order $1/S$, where
$S$ is the size of spin, the 
Holstein-Primakoff transformation for the spin operators can be written as
the following : 
\begin{eqnarray}
S_i^z = S - a^{\dagger}a, S_i^- = \sqrt{2S}a^{\dagger}(1-a^{\dagger}a/(2S))^
{1/2}
\simeq \sqrt{2S}a^{\dagger}; i \in odd\nonumber\\
S_i^z = -S +b^{\dagger}b, S_i^- = \sqrt{2S}(1-b^{\dagger}b/(2S))^{1/2}b
\simeq \sqrt{2S}b; i \in even.
\end{eqnarray}
Using this transformation, the Hamiltonian in Eq. (\ref{hamiltonian}) can be 
separated into three terms as follows :
\begin{equation}
H = H_{odd-even} + H_{odd-odd} + H_{even-even}\nonumber,
\end{equation}
where $H_{odd-even}, H_{odd-odd}$, and $H_{odd-even}$ are defined as :
\begin{eqnarray}
H_{odd-even} &=& \sum_{i,j}^L J_{2i-1,2j} [-S^2 + S(a_{2i-1}^{\dagger}a_{2i-1} 
+ b_{2j}^{\dagger}b_{2j} + a_{2i-1}b_{2j} + 
a_{2i-1}^{\dagger}b_{2j}^{\dagger})]\nonumber\\
H_{odd-odd} &=& -\sum_{i<j}^L J_{2i-1,2j-1}[-S^2 + S(a_{2i-1}^{\dagger}a_{2i-1}
 +a_{2j-1}^{\dagger}a_{2j-1} + a_{2i-1}a_{2j-1}^{\dagger} + 
a_{2i-1}^{\dagger}a_{2j-1})]\nonumber\\
H_{even-even} &=& -\sum_{i<j}^L J_{2i,2j}[-S^2 + S(b_{2i}^{\dagger}b_{2i} +
b_{2j}^{\dagger}b_{2j} + b_{2i}b_{2j}^{\dagger} + 
b_{2i}^{\dagger}b_{2j})].
\end{eqnarray}
We diagonalize this quadratic Hamiltonian
by going to momentum space and then diagonalizing by a Bogoliubov
transformation:
\begin{equation}
\label{hamiltonian_k}
H = constant +JS\sum_k\Big[ (\alpha-f(k)) (a_k^{\dagger}a_k + 
b_k^{\dagger}b_k) + g(k) (a_k^{\dagger}b_k^{\dagger}+ a_k b_k)\Big]
\end{equation}
where
\begin{eqnarray}
\label{sum}
\alpha &=& 2\lim_{L\to\infty}\sum_{n=1}^{L/2} \frac{1}{(2n-1)^{\beta}},
\nonumber\\
f(k) &=& 4\lim_{L\to\infty}\sum_{n=1}^{L/2}\frac{1}{(2n)^{\beta}}[\cos(2nk)-1],
\nonumber\\
g(k) &=& 2\lim_{L\to\infty}\sum_{n=1}^{L/2}\frac{1}{(2n-1)^{\beta}}\cos(2n-1)k;
\end{eqnarray}
using the Bogoliubov transformation, the Hamiltonian  
(\ref{hamiltonian_k}) can be diagonalized and be written in terms of 
free boson operators $c_k$ and $d_k$:
\begin{equation}
H = constant + JS\sum_k\omega_k (c_k^{\dagger}c_k + 
d_k^{\dagger}d_k)
\end{equation}
where
\begin{equation}
\label{omega}
\omega_k = \sqrt{(\alpha-f(k))^2 -(g(k))^2}.
\end{equation}
The correction to staggered magnetization is given by :
\begin{equation}
\Delta m = \frac{1}{V}\sum_k<a_k^{\dagger}a_k> = \Delta m_q + \Delta m_T(T),
\end{equation}
where $\Delta m_q$ and $\Delta m_T(T)$, which represent the 
quantum and thermal
fluctuation correction respectively, are given by
\begin{eqnarray}
\label{correction}
\Delta m_q &=& \int{\frac{dk}{2\pi}\frac{1}{2}\Big[\frac{\alpha-f(k)}
{\omega_k}}-1\Big]\nonumber\\
\Delta m_T(T) &=& \int{\frac{dk}{2\pi}
\Big[\frac{\alpha-f(k)}{\omega_k}}\Big]\frac{1}{e^{E_k/k_BT}-1}.
\end{eqnarray}
We will visit these equations frequently when we discuss the validity of the
spin wave approach later in the text. 

It is clear that the correction to magnetization is dominated by the small
$k$ behavior of the spin wave spectrum. We thus need to obtain the small $k$
behavior of the expressions
given in Eq. (\ref{sum}). To do that we express them
in terms of the Bose-Einstein integral 
function \cite{robinson} defined as :
\begin{equation}
\label{be}
F(\alpha,v) = \frac{1}{\Gamma(\alpha)}\int{dx \frac{x^{\alpha-1}}{e^{x+v}-1}}
= \frac{e^{-v}}{1^{\alpha}}+\frac{e^{-2v}}{2^{\alpha}}+ 
\frac{e^{-3v}}{3^{\alpha}}+\dots 
=\sum_{n=1}^{\infty}\frac{e^{-nv}}{n^{\alpha}},
\end{equation}
and rewrite
the $\cos(nk)$ term in $f(k)$ and $g(k)$ as the following :
\begin{equation}
\label{cos}
\sum_{n}^{\infty}\frac{\cos(nk)}{n^{\beta}} = \Re \Big[\sum_{n} \frac{e^{ink}}
{n^{\beta}}\Big] = \Re \Big[F(\beta,-ik)\Big].
\end{equation}
The analytical properties of $F(\alpha, v)$ near $v=0$ are known and are given 
by :
\begin{eqnarray}
\label{analytic}
F(\alpha,v) &=& \Gamma(1-\alpha)v^{\alpha-1} + \sum_{n=0}^{\infty}
\frac{\zeta(\alpha-n)}{n!} (-v)^n, (\alpha\notin \mathbb{Z})\nonumber\\
F(\alpha,v) &=&\frac{(-v)^{\alpha-1}}{(\alpha-1)!}\Big[\sum_{r=1}^{\alpha-1}
\frac{1}{r} - ln(v)\Big] + \sum_{n\ne\alpha-1}\frac{\zeta(\alpha-n)}{n!}
(-v)^n, (\alpha\in \mathbb{Z})
\end{eqnarray}
where $\zeta(s)$ is the zeta function.
We will use these properties in our later treatment.

\section{Spin Wave Spectra and Corrections to Staggered 
Magnetization}

In this section we analyze 
Eq.(\ref{sum}) for different values of $\beta$ to obtain the spin wave 
spectra and calculate the correction to staggered magnetization, to determine
the validity of the spin wave approach. 

\subsection{$\beta\ge3$}

Equations (\ref{cos}) and (\ref{analytic}) are the main ingredients to
analyze Equation (\ref{sum}) which can be summed up in closed forms. 
Up to leading order in $k$ the relations in Eq.(\ref{sum}) for $\beta > 3$ 
read:
\begin{equation}
\alpha = 2\sum_{n=1}^{\infty} \frac{1}{(2n-1)^{\beta}} = 
2(1-2^{-\beta})\zeta(\beta)\nonumber
\end{equation}
\begin{eqnarray}
\label{b>1}
f(k) &=& 4\sum_{n=1}^{\infty}\frac{1}{(2n)^{\beta}}[\cos(2nk)-1]\nonumber\\
&=& 2^{2-\beta}\Big[ \Re (F(\beta,-2ik)) - \zeta(\beta)\Big]\nonumber\\
&\simeq& c k^2\nonumber\\
g(k) &=& 2\sum_{n=1}^{\infty}\frac{1}{(2n-1)^{\beta}} \cos(2n-1)k \nonumber\\
&=& 2\sum_{n=1}^{\infty} \Big[\frac{\cos(nk)}{n^{\beta}}
- \frac{\cos(2nk)}{(2n)^{\beta}}\Big]\nonumber\\
&\simeq& \alpha - c'k^2,
\end{eqnarray}
where $c$ and $c'$ are positive constants.
The same results can also be obtained by expanding the $\cos(nk)$ term to order
$k^2$ in 
$f(k)$:
\begin{equation}
\sum_n{\frac{\cos(nk)-1}{n^\beta} \simeq -k^2\sum_n{n^{2-\beta}}},
\end{equation}
in which the sum converges as long as $\beta>3$; together with
a similar expansion for $g(k)$ one reproduces Eq. (\ref{b>1}).
The spin wave spectrum can be easily shown to be linear in $k$:
$\omega_k \propto k$,
and the $T=0$ correction to the staggered magnetization from long-wave length 
spin wave fluctuation:
\begin{equation}
\Delta m_q \sim \int\frac{dk}{\omega_k}  
\end{equation}
diverges logarithmically for $\beta > 3$. This immediately indicates that
the spin wave approach is not valid for $\beta > 3$ at zero temperature. 
The results obtained here are essentially the same as
the spin wave calculation for 
nearest-neighbor interactions only.\cite{anderson} 

For $\beta=3$ the expansion we did above is no longer valid because the sum
is divergent. We rely instead on the Bose-Einstein integral function
as defined in Eq. (\ref{be}) to calculate $\omega_k$ and $\Delta m_q$. 
After a little algebra we find 
$\omega_k\sim k\sqrt{|\log(k)|}$ which leads to the correction of staggered 
magnetization that diverges as $\sqrt{|\log(L)|}$, 
where $L$ is the system size.
We thus conclude that the quantum fluctuation destroys Neel order, and the 
spin wave approach is not valid for $\beta\ge3$.

\subsection{$1<\beta<3$}
We now turn our attention to the case $1<\beta<3$. 
As in $\beta=3$ case we are no 
longer able to expand the $\cos(nk)$ term in $f(k)$ and $g(k)$ 
because the coefficient of $k^2$ is divergent so we again take advantage
on the mapping onto the Bose-Einstein integral function.
In the long wave-length regime, the relations given in Eq. (\ref{sum}) read :
\begin{equation}
\alpha = 2\sum_{n=1}^{\infty} \frac{1}{(2n-1)^{\beta}} = 
2(1-2^{-\beta})\zeta(\beta)\nonumber
\end{equation}
\begin{eqnarray}
f(k) &=& 4\sum_{n=1}^{\infty}\frac{1}{(2n)^{\beta}}[\cos(2nk)-1]\nonumber\\
&=& 2^{2-\beta}\Big[ \Re (F(\beta,-2ik)) - \zeta(\beta)\Big]\nonumber\\
&\simeq&-\phi(\beta) k^{\beta-1}\nonumber
\end{eqnarray}
\begin{eqnarray}
g(k) &=& 2\sum_{n=1}^{\infty}\frac{1}{(2n-1)^{\beta}} \cos(2n-1)k \nonumber\\
&=& 2\sum_{n=1}^{\infty} \Big[\frac{\cos(nk)}{n^{\beta}}
- \frac{\cos(2nk)}{(2n)^{\beta}}\Big]\nonumber\\
&\simeq& \alpha - \frac{1}{2}\phi(\beta)k^{\beta-1},
\end{eqnarray}
where the function $\phi(\beta)$ is given by :
\begin{equation}
\phi(\beta) = \frac{\pi}{\Gamma(\beta)} \frac{1}{\cos[\pi(\beta-2)/2]},
\end{equation}
with $\Gamma(\beta)$ being the gamma function. 
The long wave-length spin wave spectrum is given by :
\begin{equation}
\omega_k \simeq 
\sqrt{3\alpha\phi(\beta)}k^{(\beta-1)/2},
\end{equation}
which is {\em sublinear},
and $T=0$ correction to staggered magnetization by :
\begin{equation}
\label{magnetization}
\Delta m_q 
\simeq \frac{1}{2\pi}\Big[\sqrt{\frac{\alpha(\beta)}{3\phi(\beta)}}
\frac{2}{3-\beta} 
\pi^{(3-\beta)/2} + \sqrt{\frac{\phi(\beta)}{3\alpha(\beta)}}\frac{2}{1+\beta}
\pi^{(1+\beta)/2}  -\pi \Big] 
%{\beta\notin \mathbb{Z}}.
\end{equation}
which is {\em convergent} for $\beta<3$. 
These results show that the system supports gapless excitations, the 
spectrum follows a sublinear power-law at
small momentum $k$, and that the Neel order at zero temperature survives, for 
large enough $S$, for $1<\beta<3$. Our results agree with an earlier work 
presented by Parreira, Bolina, and Perez \cite{parreira}who show the 
existence of Neel order for $\beta\ge3$ and the presence of Neel order for 
$\beta<3$ at zero temperature using rigorous proof. However the excitation 
spectra were not studied in this work, nor was the critical value of $S$ for 
the
stability for Neel order calculated. Another support for our results at zero
temperature is offered by the work of 
Aoki \cite{aoki} who studied the same model we are studying for the case 
$\beta=2$ in 1D and 2D using spin wave theory. 
In that work he found that there exists Neel order at zero temperature in 1 
dimension for $\beta=2$ which is in agreement with our conclusion. 

We may also estimate the critical size of
the spin, $S_c$, above which the Neel order survives, by setting the correction
to the staggered magnetization equal to the spin size: $\Delta m_q=S_c$. 
As $\beta\to3$, $\Delta m_q$ is dominated by long-wave length spin-wave 
fluctuations, and we obtain
\begin{equation}
S_c(\beta) \simeq \frac{1}{2\pi}\Big[\sqrt{\frac{\alpha(\beta)}{3\phi(\beta)}}
\frac{2}{3-\beta} \pi^{(3-\beta)/2} + \sqrt{\frac{\phi(\beta)}{3\alpha(\beta)}}
\frac{2}{1+\beta} \pi^{(1+\beta)/2}  -\pi \Big] 
\simeq \frac{0.41}{\sqrt{3-\beta}},
\end{equation}
a result we expect to be asymptotically exact in the limit $\beta\to3$.
On the other hand we also find the quantum correction gets suppressed very 
rapidly as $\beta$ decreases from 3; for examples we find
$S_c \simeq 1/2$, for $\beta =2.63$ 
and $S_c \simeq 1$ for $\beta=2.85$,
suggesting the Neel order would survive for any spin for $\beta \lesssim 2.6$.

We also calculate the correction to staggered magnetization at finite
temperature. First we discuss the case for $\beta>2$. The thermal correction 
to staggered magnetization is given by :
\begin{equation}
\Delta m_T(T) \simeq \frac{k_BT}{2\pi JS}\int{dk\Big[
\frac{k^{1-\beta}}{3\phi(\beta)} + \frac{1}{3\alpha}\Big]},
\end{equation}
which diverges as $L^{\beta-2}$ for $\beta>2$. 
For $\beta=2$, it is a simple exercise to show that the spectrum behaves like
$\omega_k \sim \sqrt{k}$, and the
small $k$ contribution to the thermal correction 
of staggered magnetization diverges as $\sqrt{|\log(L)|}$. These results
indicate 
that thermal fluctuations destroy the Neel order for $\beta\ge2$ at any 
finite temperature. They are consistent with an extension of the Mermin-Wagner
theorem that Bruno advanced,\cite{bruno} which proves the absence of Neel 
order for $\beta\ge 2$.
For classical antiferromagnets in 1D, it has been shown, using Monte Carlo 
simulation, that  there is no magnetic ordering at finite temperature. 
\cite{romano2}

For $\beta<2$ the correction to staggered magnetization is given by :
\begin{equation}
\label{finite}
\Delta m_T(T)  
\simeq \frac{k_B T}{\pi JS}\Big[\frac{\pi^{(2-\beta)}}{3(2-\beta)\phi(\beta)}
+ \frac{\pi}{3\alpha}\Big]
%; \beta \notin \mathbb{Z}.
\end{equation}
This {\em convergent} correction
shows that the Neel order survives at finite temperature for 
$\beta<2$.
The Neel transition temperature $T_N$ can also be estimated by applying the 
same
rationale used to estimate the critical value of $S$ at zero temperature. By
using Eq. (\ref{finite}) we find :
\begin{equation}
T_N(S,\beta) = \frac{\pi JS}{k_B} 
\Big[\frac{\pi^{2-\beta}}{3(2-\beta)\phi(\beta)}
+ \frac{\pi}{3\alpha} \Big]^{-1}.
\end{equation}
In the limit $\beta\to2$, 
we find $T_N$ vanishes linearly:
\begin{equation}
T_N \simeq \frac{3\pi^2 JS}{k_B} (2-\beta)
\end{equation}

We see that increasing the range of 
interactions (or decreasing $\beta$)
in the chains has effects that are
similar to increasing the dimensionality of the
systems. For $\beta\ge 3$ we find absence of Neel order at both zero and 
finite 
temperature, a genuine one-dimensional (1D) behavior.
For $2 \le \beta < 3$ we have finite Neel order at zero temperature which
gets destroyed at any
finite temperature, similar to the 2D situation. 
Finally for $\beta<2$ the Neel order is
stable at zero and low-enough finite temperature, a behavior expected for 
dimensions above two. 

In contrast to the antiferromagnetic case we are studying here, the 
ferromagnetic models with long range interactions have been studied more 
extensively. Classical Heisenberg model with long range ferromagnetic 
interactions has a phase transition at finite temperature in 1 dimension
when the interactions decay slower than $1/r^2$. There is no phase transition
at finite temperature when the interactions decay faster than $1/r^2$.
\cite{frohlich} This result
for the classical case in 1 dimension is confirmed by Monte Carlo simulation. 
\cite{romano}
The quantum Heisenberg model with long range interactions have also been 
studied using the modified spin wave theory. \cite{nakano1, nakano2} 
It was shown that there exists a magnetic 
ordering in 1 dimension as long as the interactions decay slower than $1/r^2$. 

%A similar behavior can also be found in 1D ferromagnetic system. This system
%been studied more extensively than the antiferromagnetic counterpart. 
%It has been shown that there exists an 
%ordering transition in long-range ferromagnetic model in 1 dimension 
%at finite temperature for $\beta<2$. \cite{frohlich,nakano1,nakano2}

\subsection{$\beta\le1$}

In this section we consider the case $\beta\le 1$. The reason we 
separate $\beta\le1$ case with the rest is because there are divergences in the
thermodynamic limit which
require special care in their analysis. Physically, this is closely related to
the fact that the ground state energy grows faster than the system size
(i.e., it becomes ``superextensive"), if the local 
energy scale $J$ is not rescaled according to the system
size. For this reason we will not discuss the finite temperature (or 
thermodynamic) properties of the system, as the definition of temperature
becomes somewhat ambiguous; we will focus instead on the ground state 
properties of the system, which is free of such ambiguity.

For the reasons mentioned we need to work explicitly with a finite 
system size $L$, defined as the number of spins per sublattice (so the total
number of spins is $2L$), and treat $k$ and $L$ as two independent variables. 
For a start, the summation in $\alpha$ 
\begin{equation}
\label{b<1}
\alpha = 2\sum_{n=1}^{L/2}\frac{1}{(2n-1)^\beta},
\end{equation}
diverges for $\beta\le1$ if we run the summation to infinity. For large but
finite $L$, we have
\begin{displaymath}
\alpha \simeq 
\left\{ 
\begin{array}{lcc}
\log(L) & & \beta=1;\\\\
L^{1-\beta}/(1-\beta) & & \beta<1.
\end{array}
\right.
\end{displaymath}
Similarly
\begin{equation}
f(k) = 4\sum_{n=1}^{L/2}\frac{\cos(2nk)-1}{(2n)^\beta}
\end{equation}
\begin{displaymath}
f(k)\simeq 
\left\{ 
\begin{array}{lcc}
-2(\log(k) + \log(L)) & & \beta=1,\\\\
2\Gamma(1-\beta)\cos(\pi(\beta-1)/2)k^{\beta-1} - 2 L^{1-\beta}/(1-\beta)
& & \beta<1;
\end{array}
\right.
\end{displaymath}
and
\begin{equation}
g(k) = 2 \sum_{n=1}^{L/2}\frac{\cos[(2n-1)k]}{(2n-1)^\beta}
\end{equation}
\begin{displaymath}
g(k)\simeq 
\left\{ 
\begin{array}{lcc}
\log(k) & & \beta=1\\\\
\Gamma(1-\beta)\cos(\pi(\beta-1)/2)k^{\beta-1} & & \beta<1
\end{array}
\right.
\end{displaymath}
The spin wave spectrum reads :
\begin{equation}
\label{bb=1}
E_k = JS\omega_k
= JS\alpha\sqrt{\Big(1-\frac{f(k)}{\alpha}\Big)^2 - 
\Big(\frac{g(k)}{\alpha}\Big)^2}
\end{equation}
\begin{displaymath}
E(k)\simeq 
\left\{ 
\begin{array}{lcc}
3JS\log(L)(1+b\log(k)) & & \beta=1\\\\
3JSL^{1-\beta}(1-bk^{\beta-1}) & & \beta<1
\end{array}
\right.
\end{displaymath}
which approach $L$-dependent constants as $k\to 0$.
Here $b \propto 1/\log(L)$ for $\beta=1$ and $b \propto L^{\beta-1}$ for 
$\beta<1$. Correction to staggered magnetization at zero temperature can be 
calculated easily using the relations derived above to yield :
\begin{eqnarray}
\Delta m_q&\sim& \frac{1}{\log(L)} \hspace{20pt} \beta=1 \nonumber\\
&\sim& \frac{1}{L^{1-\beta}} \hspace{20pt} \beta<1,
\end{eqnarray}
suggesting the quantum fluctuation gets completely suppressed as system size
grows.

For $\beta=0$ the calculation becomes particularly 
simple; the relations for $\alpha, f(k),$
and $g(k)$ in Eq. (\ref{sum}) become :
\begin{eqnarray}
\alpha &=& L,\nonumber\\
f(k) &=& \sum_{\boldsymbol{\delta}_{2}}\Big[e^{i\boldsymbol{k}\cdot 
\boldsymbol{\delta}_{2}} + e^{-i\boldsymbol{k}\cdot \boldsymbol{\delta}_{2}}
- 2\Big]\nonumber\\
&=& 2L(\delta_{\boldsymbol{k},0}-1),\nonumber\\
g(k) &=& \sum_{\boldsymbol{\delta}_{1}}e^{i\boldsymbol{k}
\cdot \boldsymbol{\delta}_{1}}\nonumber\\
&=& L\delta_{\boldsymbol{k},0}.
\end{eqnarray}
The spin wave spectrum for $k\ne0$ is given by :
\begin{equation}
E_k = JSL\sqrt{(1+2)^2} = 3JSL,
\end{equation}
which is $k$-{\em independent},
and the correction to staggered magnetization is given by :
\begin{equation}
\Delta m_q \sim \sum_k\frac{1}{\omega_k} \sim \frac{1}{L}.
\end{equation}
We will compare these with an exact solution for this special case in the next
section.

\subsection{$\beta=0$ : exact solution}

The infinite range ($\beta=0$)
antiferromagnetic chain with no frustration is given by the
following Hamiltonian :
\begin{equation}
H=J\sum_{ij}^{2L} (-1)^{i-j+1}\boldsymbol{S}_i\cdot\boldsymbol{S}_j,
\end{equation}
which can be solved exactly in the following manner.
We introduce:
\begin{eqnarray}
\boldsymbol{S}_{A} &=& \sum_{i\in A}\boldsymbol{S}_i,\nonumber\\
\boldsymbol{S}_{B} &=& \sum_{i\in B}\boldsymbol{S}_i,
\end{eqnarray}
where $\boldsymbol{S}_{A}(\boldsymbol{S}_{B})$ is the total spin operator 
for sublattice $A(B)$, to rewrite the Hamiltonian in the following form :
\begin{equation}
H=J\Big[\boldsymbol{S}_A\cdot\boldsymbol{S}_B - 
((\boldsymbol{S}_A^2 + \boldsymbol{S}_B^2)) +
\Big(\sum_{i\in A}(\boldsymbol{S}_i)^2 + 
\sum_{i\in B}(\boldsymbol{S}_i)^2\Big)\Big].
\end{equation}
We define the total spin operator $\boldsymbol{S}_{tot} = \boldsymbol{S}_A + 
\boldsymbol{S}_B$ to further simplify the Hamiltonian given above to become :
\begin{equation}
\label{hamiltonian_exact}
H = J\Big[\frac{1}{2}\boldsymbol{S}_{tot}^2 - \frac{3}{2}
(\boldsymbol{S}_A^2 + \boldsymbol{S}_B^2) + 3L/2\Big].
\end{equation}
The Hamiltonian in Eq. (\ref{hamiltonian_exact}) can be diagonalized in the 
total-$S$ basis of states 
given by $|(S_A,S_B);S_{tot}>$ where $S_A (S_B)$ and $S_{tot}$ are the total 
spin quantum number in sublattice $A(B)$ and in the system respectively.
Using this basis, the energy can be easily obtained as :
\begin{equation}
E = J\Big[\frac{1}{2}S_{tot}(S_{tot}+1) - \frac{3}{2}(S_A(S_A+1) + S_B(S_B+1)) 
+ 3L/2\Big].
\end{equation}
To minimize the energy we must have all spins aligned in each sublattice and
have a minimum of $S_{tot}$. This means that $S_{tot} = 0$ and $S_A = S_B 
= LS$, where $S$ is the spin size, 
will minimize the energy and give us the ground state. The momentum quantum
number of the ground state is 0 ($\pi$) for even (odd) $L$. 
The lowest energy excited state is obtained by having $S_{tot}=1$
while still maintaining maximum $S_A$ and $S_B$. The
energy gap is given by :
\begin{equation}
\Delta E = E_{ex} - E_{gs} = J.
\end{equation}
This particular excited state has a momentum quantum number that differs from
the ground state by $\pi$, which corresponds to momentum
$k=0$ in the spin wave approach, due to the doubling of the unit cell in that
approach. 
We will say more about this in the next 
section. 
To obtain excitations with generic $k$ however, we must change either the 
$S_A$ or $S_B$ quantum numbers. There exist two branches of degenerate 
low-lying excitations, corresponding to
$S_A = LS-1$ or
$S_B=LS-1$ and $S_{tot}=1$, with excitation energy 
\begin{equation}
\Delta E = E_{ex} - E_{gs} = J(1+3LS),
\end{equation}
which grows linearly with system size, and has no $k$-dependence. This result
agrees with the
spin wave solution obtained earlier 
in the limit $S\to\infty$, as expected. 

\section{Excitations at $k=0$ and Status of the Lieb-Schultz-Mattis theorem}

The Lieb-Schultz-Mattis (LSM) theorem\cite{lsm} states 
that for half-integer
spin chains with length $L$ and short-range interaction, there exist 
an excited state whose momentum differs from the ground state by $\pi$, with
energy that vanishes at least as fast as $1/L$ as $L\to\infty$. \cite{lsm} 
Recently the theorem has been extended to spin chains with power-law long
range interaction, and it was found that the theorem remains valid for 
$\beta > 2$.
\cite{parreira2,hakobyan} 
The situation is unclear for $\beta \le 2$.
 
In this section we check if the LSM behavior still holds for $\beta \le 2$, 
using the spin-wave method. As discussed above, due to the doubling of the
unit cell, excitations whose momenta differ from the ground state by either
$\pi$ or 0 show up as $k=0$ excitation in the spin-wave approach. If one 
blindly use the linear spin-wave results however, one would always find
$E_{k=0}=0$. But this is an artifact of the linear spin-wave approach 
which maps the $k=0$ modes to harmonic oscillators {\em without} a restoring 
force. Thus in order to study the excitation that are relevant to the LSM
theorem, we must treat the $k=0$ modes more carefully.

To do that, we start by rewriting the Hamiltonian as given in 
Eq. (\ref{hamiltonian}) in the momentum space:
\begin{equation}
H = \sum_k \sum_{\delta_1} J(\delta_1)\boldsymbol{S}_{k}^{A}\cdot
\boldsymbol{S}_{-k}^{B}
e^{-i\boldsymbol{k}\cdot\boldsymbol{\delta}_1} -
2\sum_{k}\sum_{\delta_2}\Big(J(\delta_2)\boldsymbol{S}_{k}^{A}\cdot
\boldsymbol{S}_{-k}^{A}
e^{-i\boldsymbol{k}\cdot\boldsymbol{\delta}_2}
+J(\delta_2)\boldsymbol{S}_{k}^{B}\cdot\boldsymbol{S}_{-k}^{B}
e^{-i\boldsymbol{k}\cdot\boldsymbol{\delta}_2}\Big),
\end{equation}
where
\begin{equation}
\label{fourier}
\boldsymbol{S}_{i}^{A/B} = \frac{1}{\sqrt{L}} \sum_k \boldsymbol{S}_{k}^{A/B}
 e^{-i\boldsymbol{k}\cdot\boldsymbol{x_i}},
\end{equation}
and $A(B)$ denotes odd(even) sublattice. Instead of applying the 
Holstein-Primakoff mapping for all terms in $H$,
we separate out the $k=0$ term in $H$ and apply Holstein-Primakoff mapping to
the $k\ne 0$ terms only. Since to linear order the $k=0$ term commutes with the
other terms in $H$, they can be diagonalized independently. The spin wave
treatment for the $k\ne0$ terms gives the spectra obtained earlier, except
that $k$ must be nonzero. On the other hand the $k=0$ term
\begin{eqnarray}
H_{k=0} &=& \frac{1}{L}\sum_{\delta_1}J(\delta_1)\Big(
\sum_{i \in A}\boldsymbol{S}_i\Big)\cdot\Big(\sum_{i \in B}
\boldsymbol{S}_i\Big)-\frac{1}{L}\sum_{\delta_2}J(\delta_2) \Big[ 
\Big(\sum_{i \in A}\boldsymbol{S}_i\Big)^2 +
\Big(\sum_{i \in A}\boldsymbol{S}_i\Big)^2\Big]\nonumber\\
&=&\frac{1}{L}\sum_{\delta_1}J(\delta_1)
\boldsymbol{S}_{A}
\cdot\boldsymbol{S}_{B} - \frac{1}{L}\sum_{\delta_2}J(\delta_2)
\Big(\boldsymbol{S}_A^2+\boldsymbol{S}_B^2\Big),
\end{eqnarray}
takes a form identical to the Hamiltonian for $\beta=0$ that was solved exactly
in the previous section.
We can easily solve this Hamiltonian to obtain the excitation energy at 
momentum $\pi$ measured from the ground state momentum, or $k=0$ for the 
doubled unit cell:
\begin{equation}
\Delta E = \frac{J\alpha}{L},
\end{equation}
where $\alpha$ depends on the power law exponent 
$\beta$ and is given by Eq. 
(\ref{sum}). For $\beta>1$, $\alpha$ is convergent in the large $L$ limit
and is given by Eq. (\ref{b>1}).
This means that the energy of the excited state vanishes as $1/L$ as 
$L\to\infty$. For $\beta=1$, $\alpha$ diverges as $ln(L)$ as shown in Eq. 
(\ref{b<1}) and the energy vanishes as $\log(L)/L$. For $\beta<1$, $\alpha$ 
diverges as $L^{1-\beta}$ as shown again in Eq. (\ref{b<1}) and 
the excitation energy vanishes as $L^{-\beta}$. 
We thus find that the LSM behavior holds for $1 < \beta \le 2$, despite the the
absence of a proof for this range of $\beta$. On the other hand the 
LSM theorem is ``violated" for $\beta\le 1$.

\section{Summary}
We have studied antiferromagnetic chain with unfrustrated long range 
interaction using the spin wave technique. We find that this approach is valid
for $\beta<3$ at zero temperature for sufficiently large size of spin,
and $\beta<2$ for sufficiently low
finite temperature, due to the stability of Neel order. 
Within the range of validity of this approach we find that the system has
a gapless excitation and the excitation spectrum follows a non trivial $k$
dependence. We also study how the excitation gap closes in this system in the 
limit $L\to\infty$, and find a behavior that is in contrast to that predicted
by Lieb-Schultz-Mattis theorem for chains with short range interactions,
when $\beta \le 1$.

%{\em Note Added}: while this manuscript was being completed, we learned of an
%earlier work\cite{parreira} which proves the absence of Neel order for $\beta 
%\ge 3$, and presence of Neel order for $\beta
%< 3$, for sufficiently large $S$, at zero temperature. Our results are in
%agreement with those of Ref. \onlinecite{parreira}. However the excitation 
%spectra were not studied in this work, nor was the critical value of S for the
%stability for Neel order calculated.

\acknowledgments
One of us (KY) thanks the Max Planck Institute for Physics
of Complex Systems in Dresden and Aspen Center for Physics for hospitality,
where parts of this work were performed, and Ian Affleck for helpful
discussions. This work was supported by NSF grant No. DMR-0225698 (EY and KY),
and the State of Florida (AJ).

\end{document}